\documentclass[12pt,letterpaper]{article}
\begin{document}

\newcommand{\be}{\begin{equation}}
\newcommand{\ee}{\end{equation}}
\newcommand{\erfc}{~{\rm erfc}}
\newcommand{\erf}{~{\rm erf}}

\author{H.~Neuberger\\ [7mm]
  {\normalsize\it Department of Physics and Astronomy, Rutgers University}\\
  {\normalsize\it Piscataway, NJ 08855, U.S.A} }

\title{The ``hard'' problem of strong of interactions.\\
{\scriptsize May 9, 2010, Yossef Dothan Memorial Lecture, Tel Aviv University}\\}

\date{}

\maketitle

\abstract{This is a write-up of a lecture at the level of a physics colloquium. 
There exists an idealized mathematical formulation of strong interactions which
has no free parameters but is known to describe the real world quite accurately.
Over the last three decades the problem has been managed with increasing success. 
An overview of some facts and a little fiction will be presented, but the question
whether the problem can now be considered ``easy'' will be left unanswered.}

\section{Introduction}

I am grateful for the honor to deliver this lecture. 
I received all my professional physics
education in this department and the preparation of this talk 
made me aware of the influence this background has had on
my research trajectory since graduation in 1979. 

Joe (Yossef), my Ph. D. advisor, stood out among the Tel Aviv particle 
theorists of my time as a student: he spoke in complete sentences using proper grammar
and enunciation. Also, unconventionally, he put content into every sentence. 
He did this naturally, without being overbearing. 
Here, I note two of his
observations: (a) Physicists, as opposed to psychologists, know their variables. (b) A track record of 
``luck'' should be kept, as representing features we do not understand.

As you will see, when it comes to strong interactions we do not always know the right variables, 
and, we need ``luck'' to find them.

\section{The practice of physics}

Using observation, guess rules. Then, apply the rules to real and imagined situations.
Check that the results are consistent with other rules 
and agree with new or existing observation. Eventually, some rules become established beyond
reasonable doubt. These rules are never an ultimate truth, but must be well defined in an
idealized framework. 

A special situation may occur when 
well established rules are known to produce certain
broad pervasive phenomena, but it is unknown how to show this from 
the rules directly, and/or 
make quantitative predictions about these phenomena. 
These are ``hard'' problems, and
the theory of strong interactions presents us with such. Other examples, 
like high $T_c$ superconductivity or turbulence exist outside particle physics.

Today, I'll be talking about the ``hard'' problem of strong interactions, whose rules have been
well established for 35 years by now. 

\section{Strong interactions}

There are four major established force types in Nature and they all are associated with local
symmetries: gravitational, weak, electromagnetic and strong. The best understood among them
is the electromagnetic force. The least understood is the gravitational force, mainly because
it is always attractive and this leads to instabilities. The LHC is hoped to clarify the nature
of the weak force. The strong force is a spoiler, since anything one sees at the LHC is 
contaminated by strong force effects. Nowadays these effects are ``managed'', but not cured. 
Two main properties of strong interactions need to be brought under control before 
achieving a cure: Chiral symmetry breaking and confinement. 

A simplification of the strong interaction 
rules as they occur in Nature is usually employed:
in it one ignores the ``heavy'' quarks and 
sets the masses of the ``light'' quarks, up, down 
and strange to zero. In addition one ignores 
all other forces, but the strong one. This 
produces a theory free of any parameters. 
Its rules must produce pure numbers quantifying 
a wide range of microscopic phenomena, among which chiral symmetry breaking
and confinement are fundamental. 

The rules are determined by the gauge symmetry group $SU(N_c)$ (of color) ($N_c=3$), 
by the representation content
of the quark fields ($\tt N_c$,$\tt \overline N_c$) and by their number (of flavors), $(N_f=3)$. The rules are most succinctly summarized by a
Lagrangian, which is the integrand of $S$, the action:
\be
{\cal S} = \frac{1}{4g^2} \int d^4 x {\rm Tr} F^2_{\mu\nu} (x) +\int d^4 x 
\overline\psi (x) \gamma_\mu [\partial_\mu - i A_\mu (x) ] \psi (x) 
\ee

$g$ is a device for expanding various predictions in 
the inverse of the logarithm of the overall scale of the process; $g$ is not a true parameter. The integrals are over four dimensional space-time.
The fields $\overline\psi$ and $\psi$ represent Dirac spin $1/2$ 
particles which carry two
types of indices (quantum numbers) one flavor, running from $1$ to $N_f$ and the other
color, running from $1$ to $N_c$. The field $A_\mu$ represents spin $1$ particles 
carrying a color index that takes values from $1$ to $N_c^2 -1$. These
indices are repackaged by making $A_\mu$ a hermitian traceless $N_c\times N_c$ matrix.
As Dirac spinors, $\overline\psi,\psi$ each have a spinorial fourfold valued index, acted
on by the four $4\times 4$ matrices $\gamma_\mu$. These matrices are Clebsch-Gordan
coefficients coupling the two spin 1/2 fields to a spin 1. The chromatic field strength 
$F_{\mu\nu}$ is a traceless hermitian matrix, antisymmetric in $\mu,\nu$, defined by:
\be
F_{\mu\nu} = \partial_\mu A_\nu - \partial_\nu A_\mu + i[A_\mu , A_\nu ]
\ee
${\cal S}$ is invariant under the change of variables $\partial_\mu-i A_\mu (x) \to h(x)
[\partial_\mu-i A_\mu (x)] h^\dagger(x)$, with $h(x)\in SU(N_c)$, a property called ``local gauge invariance''. 

\section{Hamiltonian and path integrals}

The entire Lagrangian is
fixed by the requirements of symmetry and minimalism. 
One keeps 
a minimal number of derivatives and minimal nonlinearity. 

The Lagrangian can be used to construct a Hamiltonian slightly generalizing the 
prescription one learns in Classical Mechanics courses (I was taught CM by Joe) 
to situations where the standard
definition of conjugate momenta does no produce independent variables. The classical
Hamiltonian is then quantized. 

A full definition of the theory requires an intermediate
step at which the number of degrees of freedom is made finite. There are two types of infinities to control: the fact that $x$ can run off to infinity (IR) and the fact 
that there is an infinite number of $x$-values in the smallest volume of space (UV). 
The control of the IR infinity is relatively simple. 
The control of the UV infinity is more subtle,
requiring the regulating parameter to be taken to infinity while adjusting the operators
in an elaborate, but well understood way. 

Instead of the Hamiltonian framework one can use path integral quantization. 
Then the Lagrangian itself is used,
and one defines an integral over all fields of the exponent 
of the action. Matters are helped
by a trick, which takes time to the imaginary axis, and 
replaces the Lorentz group $O(3,1)$
by the euclidean group $O(4)$. One still needs to take 
care of IR and UV limits as before.
The advantage of this formulation is that it offers a 
different way of understanding the gross
features of the physics: one searches for ``important'' 
field configurations, which dominate the
path integral for one type of observable or another. 
There is no such thing as a fermion field configuration. 
The integral
over the fermion fields isn't a generalization 
of an ordinary Riemann integral, 
unlike the integral over the $A_\mu$ fields. 
Rather, the integral over the fermions is defined by first
keeping the $A_\mu$'s fixed, and then it gives:
\be
\det [ \gamma_\mu (\partial_\mu - i A_\mu ) ]^{N_f} 
\equiv \det [G_f^{-1} (A) ]^{N_f}
\ee
Moments of 
fermion operators are defined in terms of the propagator
$G_f (A)$. 

Thus, as far as the fermions go only the inverse of an  $A_\mu$-dependent
first order partial differential operator comes in: 
\be
G_f(A)=\frac{1}{\gamma_\mu (\partial_\mu - i A_\mu ) }
\ee

This is a consequence of the
fermions entering only bilinearly in the Lagrangian, a special feature of four spacetime
dimensions and the requirement of minimality.

\section{Large $N_c$}

A further special simplification is made by setting $N_c g^2=\lambda$ and 
taking $N_c\to\infty$ at fixed $\lambda$ and $N_f$. 
Now $\lambda$ will generate a scale. 
This large $N_c$ limit  (t' Hooft's~\cite{thooft}) is known to continue exhibiting spontaneous chiral symmetry
breaking and confinement and in general to provide good approximations to a large number of
observables. The number of $A_\mu$-type degrees of freedom grows as $N_c^2$, while
that of $\psi$-type degrees of freedom grows only like $N_c$. The net
result is that fermions are relegated to the role of spectators which means that the fermion determinant can be set to 1, 
and the ``energy'' cost of an $A_\mu$-configuration 
is given by the classical action. (The ``importance'' of a configuration also
depends on an ``entropy'', 
in addition to the observable of concern.)

When the theory is regulated in  the UV and IR, 
one discovers that the series in $\lambda$ has a 
finite radius of convergence~\cite{KNN}.
A similar expansion at finite $N_c$, in $g$, is only asymptotic. The
Feynman diagrams representing the expansion in $\lambda$ are planar while for an expansion
in $g$ there is no such restriction. 
The number of planar diagrams at fixed order only
grows exponentially with the order, while the number of all diagrams at fixed order grows factorially and
this is the reason for the distinct nature of the two series. 
The planar structure of diagrams allows one to draw several qualitative conclusions,  
if one assumes confinement. These conclusions are nicely born out by experiment. In short, a
solution of the theory in the planar limit would take us a long way toward cracking the
hard problem of strong interactions.

Increasing numerical evidence coming from lattice gauge theory 
indicates that at large distances the planar theory behaves 
like a theory of weakly self-interacting strings~\cite{teper}. 
The strings do not interact with each other. 
For strings that touch, theoretically, one
guesses that the closed
string coupling constant behaves as $1/N_c^2$, while the open string constant goes as $1/N_c$.
So far, this seems to work well for long distances. Still, it is doubtful that a useful 
string description exists for short distances.

Nevertheless, things look better 
in this respect nowadays~\cite{aharony}. It seems possible to find 
descriptions of $N_c=\infty$ strong interactions at 
large distances by noninteracting string-like states.
These strings propagate on weakly curved higher dimensional spaces.
Speculating that one can make a firm theory for these strings also on
strongly curved spaces, one dreams of an ``in principle'' string description of planar
Feynman diagrams within their regime of convergence. 

\section{Chiral symmetry}

The four $\gamma_\mu$ matrices are all off diagonal in terms of $2\times 2$ blocks: 
the 
matrix 
$\gamma_5=\pmatrix{0&1\cr 1& 0}$
anticommutes with all $\gamma_\mu$.
Accordingly, the four components of $\overline\psi$ and $\psi$ separately split
into two two-components spinors each, $\overline\psi_{R,L}$ and $\psi_{R,L}$.
Only $\overline\psi_R,\psi_R$ and $\overline\psi_L,\psi_L$ are directly coupled in the 
Lagrangian. In Hamiltonian language the $RL$ 
split corresponds to a separation 
according to helicity among the quarks and the antiquarks.
Helicity is a good quantum number only for massless particles. 
A fermion number conserving 
mass term consists of $\bar\psi\psi=\overline\psi_L\psi_R+\overline\psi_R\psi_L$. For zero mass, $RL$ independent changes of basis in flavor space are a symmetry. 
Thus the Lagrangian has a $U_R(N_f)\times U_L(N_f)$ symmetry. 
The chiral symmetry group is $SU_R(N_f)\times SU_L(N_f)$
obtained from the original classical symmetry group after setting aside the two central $U(1)$'s.

In the functional integral formulation, chiral symmetry is special when compared with other
global symmetries known in physics in that it holds even at fixed arbitrary $A$. 
So, configuration by configuration in the path integral, 
the symmetry holds. The more typical situation 
is that the symmetry holds only after averaging but
is not preserved configuration by configuration. 

Although the Hamiltonian commutes with the generators of chiral symmetry, the vacuum of
the theory is not invariant under the entire group. This fact is supported by experiment.
It turns out that the light quark masses are light enough that one can perturb in them
around the massless case. This perturbation would produce the relatively light 
observed masses of pions and kaons if we assume that chiral symmetry is spontaneously broken
at zero mass. The vacuum is preserved by the subgroup in which the action of the two
factors in $SU_R(N_f)\times SU_L(N_f)$ is identical. The complementary subset, where the
factors are conjugate of each other, is not an invariance of the vacuum. Each charge
is a space integral of a local current by Noether's theorem. Integrals of the currents with slowly varying weights, 
when acting on the vacuum, create states distinct from the vacuum. When the weights are constant the energy of the excitation is zero 
because the charges commute with the Hamiltonian. For slowly varying weights, the
energies are therefore low. 
These
states are made out of massless scalar particles, the Goldstone bosons associated with the
broken generators. 
These Goldstone bosons acquire small masses when we add $m \overline\psi\psi$ to 
the Hamiltonian with $m$ small relatively to the generated scale. 
The masses square of these would-be Goldstone bosons
are linear in the perturbation, that is the quark masses. Other mesons have larger masses
that stay finite and large when the quark masses are taken to zero. 

The operator $\overline\psi\psi$ is invariant
under the subgroup preserved by the vacuum. Under the
broken transformations the operator changes. Spontaneous breakdown in the vacuum reflects
itself in the vacuum expectation value $\langle \overline\psi\psi\rangle$ being non-zero at $m=0$. This
quantity is usually referred to as the ``condensate'' and a fundamental symptom of spontaneous 
chiral symmetry breaking is the existence of a non-zero condensate. 
The condensate in the presence of the $m\overline\psi\psi$ term is a 
nonzero function of $m$. Consider first finite $N_c$,  IR ($L$) and UV ($a$) cutoffs. We assume a UV lattice cutoff, 
so $a$ denotes the lattice spacing. The IR cutoff comes in the form of a four torus of side
$L$ lattice spacings. Now the matrix $G_f$ is a finite matrix 
of size $4 L^4 N_c N_f \times 4 L^4 N_c N_f$, depending on the
gauge field and on the mass $m$. The structure in the flavor indices is a Kronecker 
delta function. 

In the path integral formalism, using continuum notation, we have
\be
\langle\overline\psi\psi\rangle \propto \int [dA_\mu] \det [ \gamma_\mu (\partial_\mu - i A_\mu ) +m ]^{N_f} e^{-\frac{1}{4g^2} \int {\rm Tr} F_{\mu\nu}^2} {\rm tr} G_f (A; m; x,x) 
\ee
If one takes the limit $m\to 0$ now the integrand vanishes, but if one first takes the
limit $L\to\infty$ and only subsequently $m\to 0$ one gets a nonzero answer. 

Let us sum over sites $x$ and divide by $L^4$. This does not change the answer
obtained after performing the $A$ integral. The summation over $x$ means that the 
object ${\rm tr} G_f (A; m; x,x)$ is replaced by ${\rm Tr} G_f(A;m)$, that is, the trace
is performed also on the $x,y$ indices of the matrix $G_f$. This leads us to the eigenvalues of
$G_f(A;m)$, which are just the inverses of the eigenvalues of $\gamma_\mu (\partial_\mu -
iA_\mu) +m\equiv D(A)+m$. Since $m$ enters additively, we only need the eigenvalues
of the massless Dirac operator $iD(A)$. As $L$ increases the latter become dense, and
random with a distribution determined by the $A$-dependent weight. Since the
trace is additive on the eigenvalues we only need the single eigenvalue 
density, $\rho_L(\mu)=\rho_L(-\mu)$. This density defines the average level spacing at $\mu$. 
Extracting a factor of $L^4$, we make the normalization of $\rho$ finite in the
$L\to\infty$ limit. 
We shall get a nonzero finite condensate if the normalized 
density of eigenvalues $\mu$ goes as
a constant for $\mu \sim 0$~\cite{bankscasher}:
\be
\langle\overline\psi\psi\rangle \propto \int d\mu \rho(\mu) \left [\frac{1}{\mu+im}+\frac{1}{-\mu+im}
\right ]
\ee
One sees clearly how the integrand seems to vanish at $m=0$, but 
its limit as $m\to 0$ is nonzero if $\rho(0)$ is nonzero.

Thus, a simple spectral property of the 
random matrix $D(A)$ around zero is equivalent to a nonzero
condensate. Zero is a special point because, 
for every $A$, $\gamma_5 D(A)+ D(A)\gamma_5 =0$, 
which produces the pairing of $\pm \mu$ in the spectrum. 
$D^2(A)$, which
commutes with $\gamma_5$ for every $A$, has a twofold degeneracy;  
in spinor space $D^2(A)$ is $2\times 2$ block diagonal with one block being given
by $W(A)W^\dagger(A)$ and the other by $W^\dagger(A)W(A)$ where $W(A)$ connects the RR components
and $W^\dagger(A)$ connects the LL components. $W^\dagger W$ and $WW^\dagger$ are isospectral. (I ignore zero modes and topology.)

The simplest random matrix probability distribution of a complex matrix $W$, otherwise
unrestricted, provides exactly the type of spectral density required for spontaneous chiral symmetry breaking~\cite{verbashur}. 
The conclusion is that about any sort of strong enough disorder induced by the
fluctuations in $A$ from configuration to configuration will produce a finite nonzero
condensate. Only weak disorder will thin out 
the spectrum at 0 sufficiently to produce $\rho_\infty (0)=0$ and a fully chirally invariant vacuum.

An even simpler picture emerges in the planar limit. First, the determinant factor can
be ignored. Second, one does not need to
take $L$ to infinity. 
A moderate $L$ is enough and produces the same $N_c\to\infty$ limit
as an infinite $L$ would~\cite{cont-red}. Third, at finite $L$ and $a$, as $N_c\to\infty$, 
a finite fraction of the entries of
$W(A)$ remain random numbers. 

It has taken about 30 years for this understanding of chiral symmetry breaking to develop.
Starting from noticing the connection between $\rho(0)$ and $\langle\overline\psi\psi\rangle$ 
at a formal level, it has been necessary to first UV regulate in a way which 
preserved chiral symmetry to assure that the realization of chiral symmetry
indeed is a purely IR phenomenon~\cite{overlap}, 
unrelated to the UV. Next, it took some time to realize
the relation to random matrix theory, and to show that the eigenvalue density, 
and the ordered eigenvalues themselves, 
can be renormalized in a meaningful way. Further, the simplification at infinite $N_c$ was understood only relatively recently~\cite{chisymlargen}.

\section{Confinement}

Mesons are made out of a quark and an antiquark of possibly different flavors and ``extra'' stuff.
Mesons can have a high spin, $J$, and mass $M$. Thinking about a blob of radius $R$ one
expects $J\propto R^2$. Experimentally one finds $J\propto M^2$ and hence $M\propto R$.
This indicates that the two quarks are in a linearly rising potential for $R$ large enough.
A simple explanation for this is that the quark pair is connected by a straight tube of 
constant chromatic energy, a flux tube of uniform structure along its length. 

Whether the quarks are moving and have an angular momentum or 
stationary, the flux tube should be 
the same -- so one guesses. This leads to a criterion for confinement~\cite{wilson}:
\be
{\rm Tr} W({\cal C}) \equiv \frac{1}{N_c}{\rm Tr}\langle e^{i\oint_{\cal C} A\cdot dx}\rangle
\sim e^{-\sigma {\cal A}}
\ee
Here ${\cal C}$ is a closed loop of rectangular shape $R\times T$, ${\cal A}=RT$ and
one takes $T\to\infty$ at fixed large $R$. The static potential is $V(R)\sim \sigma R$,  exhibiting confinement. 
This is a particular geometry realizing the more general statement that
the area law holds for 
any simple enough loop with a unique minimal area ${\cal A}$, 
asymptotically as the 
loop is uniformly blown up. 

For the rectangular geometry, one can think of the operator as describing the amplitude, 
in Euclidean space, of a state consisting of quark of mass $M=\infty$ 
fixed at a point and its 
antiquark situated at distance $R$ away. The amplitude would have 
been $e^{itE(R)}$ in real time, becoming $e^{-TV(R)}$ after rotating the time to the
imaginary axis and dropping the infinite factor $e^{-2 M T}$. 

If the gauge theory were abelian, the line integral in the exponent would represent the magnetic
flux through any area spanning the curve. Clearly, to get such a small average, the flux
must fluctuate strongly. A simple way to understand the area law would be to assume that
the flux is additively 
made out of small contributions, each associated with small patches
tessellating the minimal area spanning the curve and fluctuating independently of each other, 
in accordance with identical laws.

In a nonabelian theory this picture needs a reformulation: there 
is no additive nonabelian
flux, the role of a, specifically {\it minimal} spanning area 
is difficult to understand due to the nonlinearity of the nonabelian 
equations, and the trace
operation should have a role. 
Moreover, since the force carriers in the nonabelian case
carry charge, one must understand why the sources are not screened spontaneously. 
The screening effect is impossible to argue away dynamically, but the argument has
a limitation. The charge of the force carriers 
is like that of a quark antiquark pair with free color 
indices, with the trace subtracted. The group $SU(N_c)$ has elements corresponding to a
unit matrix times a phase $e^{\frac{2\pi i k}{N_c}},~k=0,1,..,N_c-1$; these ``center''  elements commute
with the entire group and make up an
abelian $Z(N_c)$ group. Quarks change when acted on by any $k\ne 0$ element by a phase and antiquarks by the opposite phase, leaving the gluons, the force carriers, 
unchanged. So, any external charges with ``$N$-ality'' 
$k\ne 0$ cannot be screened. There should be $N_c-1$ 
distinct string tensions, one for each $k$, $\sigma_k$. 

So, our abelian mechanism might work, if we keep
it restricted to the center $Z(N_c)$. 

We still need to detach the concept of random 
flux contributions from the minimal area.
The idea is to associate the flux contributions 
to other loops, which carry flux and link
the loop of our observable. Since loops only 
link in 3D, we need to adopt a Hamiltonian
picture and abandon, for the time being, 
the path integral. We imagine the Hilbert space
as made out of functionals of the space components of $A_\mu$. 
As the gauge invariance is local, one needs to define the representation content 
of each point in space. One places constraints on 
the the type of wave functionals that
are acceptable: they must obey Gauss's law, making them singlets. We now
view $W$ as an operator, acting by multiplication on the wave functionals. $W$ is gauge
invariant, so it acts within the Hilbert space. The confinement criterion is a statement
about the vacuum expectation value of the operator $W$. 
We now define another operator
associated with curves, $B_k({\cal C})$, which obeys~\cite{thooftb}:
$B_k({\cal C}) W_{k'} ({\cal C'}) = e^{\frac {2\pi i l k k'}{N_c}} W_{k'} ({\cal C'}) B_k({\cal C})$.

Here $k,k'$ denote possible $N$-alities 
and $l$ is the number of times the loops link.
$B$ creates a tube of $Z(N_c)$ magnetic flux along its curve 
producing a small patch of flux on any 
area spanning the curve of $W$ at the point where
the $B$-curve pierces it. 
Piercing has to happen if the curve of $B$ links that of $W$. 
Two loops can link and still be distant from each other and this
makes the commutation relation a nontrivial constraint in a theory describable
by a local Lagrangian. 
The commutation relation
shows that the $W,B$ are exponential versions of canonically conjugate pairs. 

If the operator $B$ condenses in the vacuum, one can imagine these
small patches to be independent of each 
other and produce confinement. Alternatively,
it may be that $W$ condenses, in which case $B$ 
would obey an area law. The operator 
that condenses, itself, cannot obey an area law if all 
particles become massive, as the energy cost would now be
concentrated at the perimeter. 
By ``condenses'' one means, roughly, that the vacuum state
is a coherent state relative to that operator. 
It is also possible to not condense neither $W$ nor 
$B$, in which case
both operators obey perimeter laws, but then
massless particles have to be present in order to
realize the commutation rule. Is is possible to 
condense both, in which case both
operators have area laws. 

The eigenvalues of the formally unitary 
matrix $e^{i\oint A\cdot dx}$ are also gauge invariant and the
action of $B$ moves them all cyclically round the circle in steps of $\frac{2\pi}{N_c}$.
For small loops we know that the nonlinearities are weak and therefore the eigenvalues
will concentrate near unity. 
Thus, if $B$ condenses, the patches of 
flux should have a typical physical scale, so that 
the eigenvalues associated with the small loop can be 
nonuniform round the unit circle. 

At infinite $N_c$ the eigenvalues freeze. 
The reason is that there are only $N_c-1$ of them
while the probability distribution governing them has an exponent that must go as $N_c^2$ for
large $N_c$. In addition, for large $N_c$ the eigenvalues form a continuum, just as before
when we were discussing the eigenvalues of the Dirac operator. 
For a small enough loop
the support of the eigenvalue distribution does not even reach round the unit circle.
For a large loop we need an almost uniform distribution, caused by the fluctuations of
the patches of flux associated with the condensation of $B$. So, we learn that at infinite
$N_c$ the world of small loops is nonanalytically different from that of large loops.
Again, the precise mechanism behind this transition as one watches a dilating loop is
not important except that it induces enough randomness among the eigenvalues. The 
condensation of $B$ is an agent causing it, but many other effects could also cause it.
After all, even with a perimeter law the eigenvalues still would have to spread out
over the entire unit circle. $B$ condensation enters only for large loops,
when one needs to explain the fact that $\sigma_k\ne 0$ for $k\ne 0$.
The transition between a gapped distribution of eigenvalues to an
ungapped one (on the unit circle) is of a generic type, likely universal. 

To understand how the strong interaction 
forces go from being weaker than electromagnetic forces
at short distances to being 
confining at long distances, we need to gain control of this
transition at infinite $N_c$. It is replaced by a 
narrow crossover at finite $N_c$~\cite{wilev}.

One problem at $N_c=3$ is that the loop-operator 
framework is defined in Hamiltonian language, 
while our basic validating technique is to
search for important configurations in the Euclidean path integral. There has 
been ongoing research on this 
for two decades. In spite of strong support
of the picture I reviewed, no overwhelming evidence has been yet found~\cite{greensite}. 

The understanding of the crossover at large $N_c$ is a relatively new activity.
It's main objective is to lay the ground for a possible future quantitative approach
to QCD at all scales, which circumvents the problem of identifying the source
of confinement by using the paradigm of matched effective theories to
connect the field theory of short distances to a string theory description of large 
distance effects.

\section{Summary}

In summary, the phenomena of spontaneous chiral symmetry 
breaking and confinement look nowadays less mysterious, but, I
for one, would not agree that the ``hard'' problem of strong interactions is already 
under control.
 
We are still looking for the right variables, and we could use some luck in our search.

\subsection*{Acknowledgments.}

I thank the sponsors of the Emilio Segre Distinguished 
Lectures in Physics of the Raymond and Beverly Sackler Foundation and the US DOE (DE-FG02-01ER41165) for support. I am very grateful for the hospitality of Lilli
Dothan, Marek Karliner and Yaron Oz and of many other members of the  
School of Physics and Astronomy at Tel Aviv University. 

Also, I wish to express my thanks to the department which turned me into a physicist some
30 years ago. In particular, in addition to Joe Dothan, the following people, among others, 
played a central role in shaping me scientifically: Yakir Aharonov, Aharon Casher, Shmuel
Nussinov and Lenny Susskind. And, I should not forget to mention the friendly cafeteria in the Physics building, where everything was happening.

\end{document}